\PassOptionsToPackage{svgnames}{xcolor}
\documentclass[sigconf,screen=true,urlbreakonhyphens=true,nonacm]{acmart}

\newif\ifeprint
\newif\ifacm
\newif\ifanonymous
\newif\ifdraft
\newif\iflipics

\eprinttrue
\acmfalse
\anonymousfalse
\draftfalse
\lipicsfalse

\eprinttrue
\acmfalse
\anonymousfalse
\draftfalse

\makeatletter
\renewcommand{\@parfont}{\bfseries}
\renewcommand{\@ACM@checkaffil}{}
\makeatother
\RequirePackage{graphicx}
\graphicspath{{./img/}}

\usepackage{amsmath,amsfonts,amsthm}
\usepackage{xspace}

\usepackage[disable]{todonotes}

\usepackage{hyperref}
\usepackage{cleveref}
\usepackage{lipsum}
\usepackage{listings}
\usepackage{xcolor}
\usepackage{tikz}
\usetikzlibrary{arrows.meta, positioning, shapes.geometric, calc, fit, backgrounds}
\usepackage{msc}
\usepackage{booktabs}
\usepackage{colortbl}
\usepackage{multirow}
\usepackage{caption}
\usepackage{graphicx}
\usepackage{pdflscape}
\usepackage{makecell}

\definecolor{TamarinBlue}{RGB}{42,0,255}
\definecolor{TamarinGreen}{RGB}{48,110,32}
\definecolor{TamarinPurple}{RGB}{175,36,67}

\lstdefinestyle{tamarin}{
    basicstyle    = \linespread{1}\footnotesize\ttfamily,
    extendedchars = true,
    tabsize       = 2,
    columns       = fixed,
    numbers       = left, 
    numbersep     = 5pt,   
    breaklines    = true,
    literate      = {~}{{\raisebox{0.5ex}{\texttildelow}}}{1},
    literate      = {>}{\textgreater}{1},
    keywords  =[1]{theory, builtins, restriction, equations, functions,
            rule, let, in, lemma, All, Ex, not, prio, deprio, presort, tactic},
    keywords  =[2]{regex},
    keywordstyle  =[1]\color{TamarinPurple},
    keywordstyle  =[2]\color{TamarinBlue},
    morecomment   = [l]{//},
    morecomment   = [s]{/*}{*/},
    commentstyle  = \color{TamarinGreen},
    xleftmargin   = 0mm,
    upquote       = true,
}

\setuptodonotes{noinline}

\newcommand{\z}[1]{\ensuremath{\mathit{#1}}\xspace}
\newcommand{\zsf}[1]{\ensuremath{\mathsf{#1}}\xspace}
\newcommand{\tvar}[1]{\z{\##1}}

\newcommand{\lightrule}{\arrayrulecolor{gray}\specialrule{0.2pt}{2pt}{2pt}\arrayrulecolor{black}}

\AtBeginDocument{%
  }

\begin{document}

\title{Less Effort, Shorter Proofs: Reinforcement Learning\\
for Security Protocol Analysis in Tamarin}


  \author{Matthias Cosler\textsuperscript{1}, Cas Cremers\textsuperscript{1}, Bernd Finkbeiner\textsuperscript{1,2}, Mohamed Ghanem\textsuperscript{1}, Niklas Medinger\textsuperscript{1}}
  \renewcommand{\shortauthors}{Cosler, Cremers, Finkbeiner, Ghanem, Medinger}
  \affiliation{%
    \textsuperscript{1}\,CISPA Helmholtz Center for Information Security, Saarbr\"ucken, Germany\\
    \textsuperscript{2}\,Technical University of Munich, Munich, Germany\\[0.3em]
    {\tt \{matthias.cosler, cremers, finkbeiner, mohamed.ghanem, niklas.medinger\}@cispa.de}
  }

\begin{abstract}

Tools like Tamarin and ProVerif have achieved notable success in analyzing and verifying
complex real-world protocols such as EMV, 5G, and WPA2, even detecting zero-day exploits. 
Despite these successes, verifying such protocols remains a time-consuming, challenging task,
often requiring significant human effort and expertise.

In this paper, we present a reinforcement learning (RL) framework inspired by AlphaZero
and AlphaProof that implements a new style of proof search for Tamarin.
We have developed a stateless API for Tamarin that acts as a classical RL environment. We guide a
Monte Carlo Tree Search (MCTS) by a neural heuristic that learns from completed subproofs.
We evaluate our framework on 16 case studies, ranging from classical protocol models to
challenging state-of-the-art protocol models from recent publications.

%
%
%
Our method finds more proofs automatically than Tamarin's standard search and produces
shorter proofs than both the standard and human-engineered heuristics.
Our pipeline is applicable out of the box to assist Tamarin users in active research, reducing the human effort required.
Moreover, our standardized interface provides a programmatic way for users to interact with Tamarin.
Finally, our work demonstrates the promising potential of adapting RL-based methods to the Tamarin domain.

\end{abstract}

\maketitle

\section{Introduction}
\label{sec:intro}
Over the last years, symbolic security protocol verification has had many success stories,
as tools like Tamarin~\cite{meier2013tamarin} and ProVerif~\cite{blanchet2001efficient} have been used to verify large, complex real-world protocols,
like WPA2~\cite{wpa2analysis}, TLS 1.3~\cite{tls130rtt,tls13symbolic},
5G~\cite{5Gformalanalysis21,5Ganalysis18}, EMV~\cite{emv2020},
Signal~\cite{kobeissi2017automated}, Apple's iMessage~\cite{linker2025formal},
SPDM~\cite{cremers2023formal,cremers2025breaking}, and WireGuard~\cite{donenfeld2017formal},
making them the de-facto standard verifiers in the field.

Despite these successes, verifying such large real-world protocols remains
a time-consuming, challenging task, often requiring significant human expertise
and manual effort.
This is due to the fact that the verifiers used often fail to scale
to the size and complexity of state-of-the-art security protocols, resulting
in non-terminating proof attempts. In this case, the
user might have to simplify the model, potentially decreasing its
faithfulness to the real protocol, and thus the value of the verification.
Alternatively, if the verifier allows it, the user has to manually assist the tool's proof effort in some way.

In the case of the Tamarin prover, users can inspect failed proof
attempts via its graphical user interface to find out where and why the
proof failed. Then they can assist the tool by providing it with a heuristic
that selects \emph{better} proof steps to apply to the proof state, or
by specifying intermediate lemmas which reduce the search space and need
to be independently proven~\cite{tamarinbook}.
Should all of these efforts fail to guide Tamarin towards an automatic
proof, users may even manually construct a proof by exploring the
search tree themselves and applying proof steps as they see fit.

All of these forms of human assistance require expert insight into the protocol being verified and the internals of the Tamarin prover to understand \emph{why} an automatic proof fails and \emph{how} to accomplish it.
Acquiring this insight is time-consuming and requires multiple iterations of trial and error, as users have to repeatedly inspect failed proof attempts. As a result, verifying large, complex protocols can take up to months~\cite{wpa2usenixsite,cremers2023formal,cremers2025breaking,5Ganalysis18}.

The human effort still required places symbolic protocol verification well behind the automation level achieved in other
domains, such as game-playing, e.g., Chess and Go, or, more recently,
mathematical theorem proving. Like symbolic protocol verification, these domains
are also characterized by large, potentially infinite search spaces and complex
decision making.

In the case of game-playing, Google's AlphaZero~\cite{silverGeneralReinforcementLearning2018} achieved superhuman
performance in Go, Chess, and Shogi by combining neural networks that evaluate the
game state and predict promising moves with a Monte Carlo Tree Search (MCTS)
algorithm that explores the game tree guided by the neural network's predictions.
AlphaZero was trained purely through self-play via reinforcement
learning, without any human data or domain-specific heuristics.

In the case of theorem proving,
AlphaProof~\cite{hubertOlympiadlevelFormalMathematical2026} recently claimed
silver-medal-level performance at the International Mathematical Olympiad (IMO) 2024
by adapting a similar combination of neural networks and MCTS to the
domain of formal mathematics in the Lean theorem prover.
Following the same principle as AlphaZero, AlphaProof generates its own training data
through proof attempts rather than self-play. Additionally, it was pretrained on mathematical text and
uses test-time reinforcement learning (TTRL) for particularly hard problems.
These achievements raise the question of whether similar machine learning
and search techniques can be applied to the domain of symbolic protocol
verification, increasing automation and reducing the human effort required.

In this work, we address and positively answer this question by
developing a reinforcement learning system that guides the proof search of the
Tamarin prover, learning from previous proof attempts to select better proof
steps and find proofs that previously could not be found without human heuristics.

To implement our RL system, we had to overcome multiple challenges:
\emph{First}, Tamarin is primarily an automated model checker. While it also offers some interactivity via a graphical user interface, there is no interface to programmatically interact with its proof search, which is required to integrate it into a reinforcement learning system.
\emph{Second}, each step in Tamarin's proof search is computationally expensive, as it invokes Tamarin's constraint-solving procedure, thus limiting the search space we can explore.
This is contrary to classical reinforcement learning environments, where the environment is typically fast to interact with, allowing for millions of interactions during training.
\emph{Third}, Tamarin is specialized for symbolic protocol verification, a domain with far fewer existing formal proofs than first-order mathematics. In contrast, related work in theorem proving relies heavily on pretraining on large corpora of existing proofs (in the case of AlphaProof~\cite{hubertOlympiadlevelFormalMathematical2026}, 80 million auto-formalized statements). This makes pretraining at a comparable scale infeasible, so our agent must learn effectively from self-generated experience on individual protocols, without access to large-scale training data.

\paragraph{Contributions.}
Our contributions are:

\begin{enumerate}
\item We extend the Tamarin prover with a reusable API, allowing users to
      programmatically interact with it.
      Using this API, Tamarin becomes a stateless REST server queried by
      external clients that store the state of the proof search. This design
      allows users to easily implement custom proof search algorithms,
      including reinforcement learning-based ones, without having to modify Tamarin's source code.
\item We use this API to implement an RL agent combining a
      tree search algorithm inspired by AlphaZero~\cite{silverGeneralReinforcementLearning2018} with a neural model that learns to interpret the proof
      state and outputs a policy, i.e., which proof method to apply next, and
      a value, estimating the distance to a complete proof from this state.
\item We evaluate our RL agent on a set of protocols from the Tamarin prover's
      GitHub repository and recent publications. We show that our agent can learn
      to prove properties in these protocols that were previously unprovable by
      Tamarin's default search strategies without human heuristics, and that it consistently finds shorter proofs than the human heuristics.
\end{enumerate}
Our code for the Tamarin API\footnote{\url{https://github.com/niklasmedinger/Tamarin-ML-API}} and the RL agent\footnote{\url{https://github.com/MatCos/TamRL}} is open-source and available on GitHub.


\paragraph{Outline.}
This paper is structured as follows: In \Cref{sec:background_ml} and \Cref{sec:background_tamarin}, we provide background on machine learning and the Tamarin prover.
We describe our reusable API for Tamarin in \Cref{sec:tamarin_env} and our reinforcement learning agent that guides Tamarin's proof search in \Cref{sec:ml}.
In~\Cref{sec:case_studies}, we present our case studies and experimental results. We present related work in \Cref{sec:related_work}, and we conclude in \Cref{sec:conclusion}.

\section{Background on Machine Learning}
\label{sec:background_ml}
Our approach uses reinforcement learning to train a neural network that guides a tree search over Tamarin's proof space.
This section gives the necessary background on these three components: Reinforcement learning provides the training framework, Monte Carlo Tree Search (MCTS) is the search algorithm that explores possible proof steps, and a Transformer network learns which proof steps are promising.

\paragraph{Reinforcement Learning.}

In reinforcement learning~\cite{suttonReinforcementLearningSecond2020}, an agent interacts with an environment in discrete steps: At each step, the agent observes the current state~$s$, selects an action~$a$, receives a scalar reward~$r$, and the environment transitions to a successor state~$s'$.
The agent's objective is to learn a \emph{policy}, a mapping from states to distributions over actions, that maximizes the expected cumulative reward.
The \emph{value function} $V(s)$ estimates this reward for state~$s$ under the current policy.
We differentiate between the \emph{state} of the environment and the agent's \emph{observation}. While the state captures all information that determines the environment's dynamics, the observation can be a partial or transformed representation of it.

\paragraph{Monte Carlo Tree Search and AlphaZero.}

Monte Carlo Tree Search (MCTS) is a best-first search algorithm that builds a search tree incrementally through repeated simulations~\cite{browne2012survey}.
Each simulation consists of four phases: \emph{selection}, where the algorithm traverses the tree from the root, choosing the most promising child at each node; \emph{expansion}, where a new node is added; \emph{evaluation}, where the new state's value is estimated; and \emph{backup}, where the obtained value is propagated back up the path to update ancestor statistics.
The classical UCT algorithm~\cite{kocsisBanditBasedMonteCarlo2006} balances exploitation and exploration by deriving a selection formula from upper confidence bounds for multi-armed bandits.
AlphaZero~\cite{silverGeneralReinforcementLearning2018} extends MCTS by replacing the random rollout with a neural network $f_\theta$ that, given a state~$s$, outputs a \emph{prior distribution} $\mathbf{p}$ assigning a probability to each action and a scalar \emph{value estimate} $v \approx V(s)$. This constitutes an actor-critic architecture: The policy head (actor) proposes promising actions, while the value head (critic) evaluates state quality.
For selection, AlphaZero uses the PUCT rule~\cite{silverMasteringGameGo2017,silverGeneralReinforcementLearning2018}, which replaces the UCT exploration term with one guided by the learned priors and values:
\[
  \text{PUCT}(s,a) \;=\;
    \underbrace{\frac{V(s,a)}{N(s,a)}}_{\text{value}}
    \;+\;
    \underbrace{c \cdot \frac{\sqrt{N(s)}}{1 + N(s,a)} \;\cdot\; p(a \mid s)}_{\text{prior}},
\]
where $V(s,a)$ is the accumulated value of the child reached via action~$a$ in state~$s$, $N(s,a)$ the visit count of action~$a$, $N(s) = \sum_b N(s,b)$ the parent's total visit count, and $p(a \mid s)$ the prior probability from the policy network.
After each completed game, the MCTS visit counts are normalized into a search policy $\boldsymbol{\pi}$, which serves as the training target for the network's prior $\mathbf{p}$, while the game outcome serves as the target for the value $v$.
The network is trained to match $\mathbf{p} \to \boldsymbol{\pi}$ via cross-entropy and $v$ to the game outcome via mean squared error, creating a positive feedback loop between search quality and network accuracy.

\paragraph{Transformers.}
The Transformer~\cite{vaswaniAttentionAllYou2017} is the dominant neural network architecture in modern deep learning.
Its core mechanism is \emph{attention}: Given a sequence of input elements, each element computes a learned compatibility score with every other element and aggregates their representations weighted by these scores.
This allows each element's representation to be informed by all other elements, with the network learning which relationships are relevant.
\emph{Multi-head attention} applies multiple independent attention operations in parallel, each capturing different aspects of the input.

A Transformer encoder stacks multiple layers of multi-head attention and feed-forward networks.
The input is first split into tokens and mapped to fixed-size vectors via a learned embedding layer.
Since attention is permutation-invariant, \emph{positional encodings} are added to provide order information.
After processing, the per-token representations can be aggregated to produce fixed-size representations for variable-length inputs, which then serve as the basis for downstream predictions such as the prior distribution and value estimate.

\section{Background on Tamarin}
\label{sec:background_tamarin}
The Tamarin prover~\cite{meier2013tamarin} is a state-of-the-art model checker for
symbolic security protocol verification. It takes as input a protocol model, a
symbolic model of cryptography, and the desired
security properties of the protocol. It then tries to find a proof or counterexample of
the properties
in the presence of a \emph{Dolev-Yao} adversary~\cite{dolev1983security} that
can intercept, modify, construct, and inject messages.

Tamarin performs \emph{unbounded} verification, i.e., there are no bounds
on the number of protocol sessions, the number of messages, the number of
agents running the protocol, and, thus, the size of a counterexample.
As a result, the problem of verifying whether a
property holds is undecidable~\cite{tiplea2005decidability,durgin2004multiset}, and
Tamarin may fail to terminate --- not finding a proof nor a counterexample.
In this case, Tamarin's users can use its graphical user interface (GUI) to
manually investigate the failing proof attempt to find out where and why
Tamarin fails to finish the proof. If they identify a reason for the failure,
they can assist Tamarin by providing a heuristic that selects \emph{better}
proof steps to use during proof search, or by specifying intermediate lemmas,
which need to be independently proven, acting as invariants and reducing the
search space.

\paragraph{Protocol Specification.}
Tamarin works in the \emph{symbolic} model of cryptography, where protocol messages
are represented as \emph{terms}, cryptographic primitives are represented as function
symbols, and an equational theory $E$ defines their semantics.

For instance, the equation $\z{adec(aenc(m, pk(sk)), sk)} = m$
models the semantics of asymmetric encryption: Decrypting a message $m$ encrypted
with the public key $pk(sk)$ using the secret key $sk$ yields the original message $m$.
Tamarin comes with built-in support for many common cryptographic primitives, but it
also allows its users to specify their own function symbols and equations,
letting them model a wide range of cryptographic primitives.
We refrain from giving a complete introduction to Tamarin's term algebra and
instead refer the reader to~\cite{tamarinbook} for more details.

To specify the protocol's semantics, Tamarin uses \emph{multiset rewriting rules} of the form
\[
[\zsf{Premises} ] \relbar\!\!\![\zsf{Actions}]\!\!\!\rightarrow [\zsf{Conclusions}].
\]
The \emph{premises} and \emph{conclusions} of a rule are multisets of \emph{facts}, which are
special or user-defined symbols applied to terms. These facts are used by the
user to model
the different protocol participants and their internal state, and by Tamarin, for instance, to model the adversary's knowledge.
The premises of a rule specify the facts that are necessary to execute
the rule, while the conclusions specify the facts that are produced by this rule.
The actions of a rule are used to label the rule with \emph{action facts} that
can then be used to specify properties of the protocol's executions via \emph{lemmas}.

\begin{example}
\label{ex:tamarin_rule}

The following rule models
the lookup of the public key \z{pkS} of a server \z{S} by a client, which then sends
an asymmetrically encrypted session key \z{k} to it:
\[
  \begin{array}{@{\hspace{10mm}}l@{\hspace{10mm}}c}
    [\ \zsf{Fr}(\z{k}),\ \zsf{!Pk}(\z{\$S},\ \z{pkS})\ ]
      \\[0.6ex]
    \relbar\!\!\![\ \zsf{SessionKey}(\z{\$S},\ \z{k})\ ]\!\!\!\rightarrow
      \\[0.6ex]
    [\ \zsf{Client\_1}(\z{\$S},\ \z{k}),\ \zsf{Out}(\z{aenc}(\z{k},\ \z{pkS}))\ ]
      \\
  \end{array}
\]
The fact \zsf{Fr}, called \emph{fresh}, is provided by a
built-in rule that models generating a unique, random value \z{k} for each
execution of the rule. This fact is commonly used to generate nonces and
cryptographic keys in Tamarin models.
The fact \zsf{!Pk} is a user-defined fact
that is used to store the public key of a server and its public identity \z{\$S}.
The ! in front of the fact indicates that it is a \emph{persistent} fact,
which, unlike linear facts, is not consumed when the rule is executed. The \zsf{Client\_1} fact is also
user-defined and stores the server's identity as well as the session key for the next
step of the client's protocol execution. Finally, the \zsf{Out} fact is a built-in
fact that models sending a message to the adversary-controlled network.

\end{example}

\paragraph{Semantics \& Protocol Properties.}
The multiset rewriting rules of a Tamarin model induce a labeled transition system,
whose state is a multiset of facts and whose transitions are labeled with the
action facts of the rules. The initial state of the system is the empty multiset.
A rule can be executed if its premises are contained in the current state, and
executing it leads to a new state where the linear premises are removed and the
conclusions are added. A sequence of rule executions starting from the initial
state is called an \emph{execution}, and the sequence of action facts of the 
rules in an execution is called the execution's \emph{trace}.

Tamarin allows the user to specify the protocol's desired properties as
\emph{trace properties} in a fragment
of first-order logic over the traces $\z{Tr}$ of a Tamarin model.
Trace properties are inductively built from atomic formulas using the usual
logical operators as well as universal and existential
quantification over terms and timepoints in the trace. The atomic formulas are
$\top$ and $\bot$, which, respectively, always and never hold,
action formulas $f@\tvar{i}$, which hold if the action fact $f$ occurs at timepoint $\tvar{i}$ in the trace,
equality of timepoints $\tvar{i} = \tvar{j}$, which holds if the timepoints
$\tvar{i}$ and $\tvar{j}$ are the
same, and equality of terms $s =_E t$, which holds if the terms $s$ and $t$ are
equal modulo the equational theory $E$.

Tamarin allows for both existential and universal trace properties, i.e., properties
that only need to hold for a single trace of the system or for all traces of the
system, respectively. Existential trace properties are usually used by users to
\emph{test} their protocol model by checking that there exists an execution that
achieves the protocol's goal, e.g., setting up a shared session key between two
parties. Universal trace properties are used to specify the security properties
a protocol should achieve, e.g., that the session key cannot be learned by
the adversary.

\begin{example}
The following universal trace property is a standard formulation of a \emph{secrecy} property.
If a session key \z{k} is established between a server \z{S} and a client, formalized by the \zsf{SessionKey(\z{S}, \z{k})} action fact, and
the adversary learns \z{k}, formalized by the built-in \zsf{K(\z{k})} action fact, then this leads to a contradiction $\bot$:

\begin{align*}
        & \forall\ t \in \z{Tr}.\ \forall\ \z{S}\ \z{k}\ \tvar{i}\ \tvar{j} . \\
        & \zsf{SessionKey}(\z{S}, \z{k})\ @\ \tvar{i}
        \wedge \zsf{K}(\z{k})\ @\ \tvar{j} \\
        & \Rightarrow \ \bot
\end{align*}

\end{example}

\section{A Reusable API for Proof Search in Tamarin}
\label{sec:tamarin_env}
In this section, we describe our first contribution: a reusable API for the Tamarin prover
that allows users to programmatically interact with its proof search. While we use this
API to implement a reinforcement learning agent in~\Cref{sec:ml}, we want to highlight that
it can be used for any proof search algorithm and thus facilitates future research on
custom proof search strategies for Tamarin, including but not limited to machine learning-based ones.


We first give an overview of Tamarin's current proof search and how users guide
it via \emph{heuristics}, and we then describe our API and how it facilitates a
new way of interacting with Tamarin via an external client that stores the state of
the proof search and only queries Tamarin to execute proof steps and validate
the final proof.

\subsection{Tamarin's Proof Search and Heuristics}
\label{par:tamarin-heuristics}

To prove a property $\psi$, Tamarin proceeds as follows:
If $\psi$ is an existential trace property, Tamarin directly searches for a trace that satisfies $\psi$, proving the property if it finds one.
If $\psi$ is a universal trace property, Tamarin uses
the duality of existential and universal quantifiers:
It proves $\psi$ for all traces by showing that there is no trace satisfying $\neg \psi$.
To show this, Tamarin exhaustively searches for a trace that satisfies $\neg \psi$ with the intention of not finding one.

This approach allows Tamarin to prove both existential and universal trace properties by searching for traces that satisfy them or their negation respectively.
During this search, Tamarin symbolically represents the candidate traces by \emph{constraint systems}.
A constraint system $\Gamma$ is a set of \emph{constraints}, which assert, for instance, that a multiset rewriting rule is executed at some timepoint in the trace, that a logical formula holds, or that the conclusion of a rule execution is consumed by the premises of another.

To search for a trace, Tamarin starts with an initial
constraint system $\Gamma^0 = \{ \psi \}$, which exactly captures the set of traces satisfying $\psi$.
Then, it recursively applies \emph{constraint solving rules} (also called \emph{proof methods}) of the form $\Gamma \rightsquigarrow \Gamma_0 \times \ldots \times \Gamma_n$, which, when applied to a constraint system $\Gamma^i$, produce new child constraint systems $\Gamma^{i+1}_0 \times \ldots \times \Gamma^{i+1}_n$. 
Since Tamarin's constraint solving rules are sound and complete, the set of traces captured by $\Gamma^i$ is exactly the union of the sets of traces captured by $\Gamma^{i+1}_0 \times \ldots \times \Gamma^{i+1}_n$.
When a rule application produces more than one child constraint system ($n > 1$), it is called a \emph{case split}.
By applying these constraint solving rules recursively, Tamarin creates a tree, refining the initial constraint system $\Gamma^0$ into more and more concrete child constraint systems.

This continues until Tamarin finds a \emph{proof}, or \emph{proof tree}, which contains a branch with a solved constraint system or only branches that end in a contradictory constraint system.
In the former case, the proof shows that a trace satisfying the initial constraint system exists, while in the latter case, the proof shows the absence of such a trace.
If the property is universal, a solution corresponds to a counterexample, and the absence of solutions corresponds to a proof. If the property is existential, a solution is a proof (witness), and the absence of solutions is a proof that the property does not hold.

\begin{example}
When proving the universal property from the previous example, Tamarin will first
negate it, resulting in the existential trace property $\phi$
\begin{align*}
        & \exists\ t \in \z{Tr}.\ \exists\ \z{S}\ \z{k}\ \tvar{i}\ \tvar{j} . \\
        & \zsf{SessionKey}(\z{S} , \z{k})\ @\ \tvar{i}
        \wedge \zsf{K}(\z{k})\ @\ \tvar{j}.
\end{align*}
Then it will simplify the initial constraint system $\Gamma^0 = \{ \phi \}$ by applying constraint solving rules that instantiate the existentially quantified variables and split the conjunction, resulting in the following constraint system
that contains two action formulas:
\[
  \Gamma = \{\ \zsf{SessionKey}(\z{S}, \z{k})\ @\ \tvar{i}, \ \zsf{K}(\z{k})\ @\ \tvar{j}\ \}.
\]
This constraint system exactly captures all traces where the adversary learns the session key
$\z{k}$ established with a server \z{S}.
\end{example}

To search for a solved constraint system, Tamarin has to make two decisions:
First, it has to decide which constraint solving rule to apply to the current
constraint system. Second, it has to decide how to explore the resulting tree of child
constraint systems. To solve the first problem, Tamarin uses heuristics,
which rank the rules $\rightsquigarrow_0,\ \ldots, \rightsquigarrow_n$ 
applicable to a constraint system $\Gamma$.
Tamarin features many inbuilt heuristics that are suited for a large class of
protocols and achieve automatic proofs on them, but users can also provide their
own heuristics via different means, as we will see shortly.

The second problem is more intricate than it first seems:
The search space might be infinite in depth, as there is no
bound on the size of the counterexample that Tamarin might find.
Thus, the tree search must not
get lost in infinite depth branches since this could cause it to miss a solved
constraint
system in a different branch. To solve this problem, Tamarin uses iterative
deepening depth first search (IDDFS) by default, but it also supports breadth
first search (BFS). Both search strategies will explore the complete search 
space up to a certain depth, and only then increase their depth bound, ensuring
that no solution is missed.

Both of these search strategies are greedy, i.e., they always use the highest ranked 
constraint solving rule according to a heuristic. This greedy search comes with
upsides and downsides: On the one hand, it can lead to fast proofs if the heuristic
is informative. On the other hand, it will lead to much longer proof
times and proof sizes or even failure if the heuristic is bad.

Should Tamarin fail to find a proof, users have to manually
investigate where in the search Tamarin fails and why this is the case:
Is Tamarin failing to prove a property which actually holds? Is it failing
to find a counterexample? Is a found counterexample a true positive, i.e., the property does not hold, or is it due to a bug in the protocol model
or property formulation that needs to be addressed?
To answer these questions, users most commonly use Tamarin's GUI to manually explore the search
tree, inspecting constraint systems and applying constraint solving rules to them.
When inspecting a constraint system, users try to identify whether the traces represented by this system are ultimately contradictory, i.e., whether the system should lead to contradictory
child systems, or whether these traces
are valid executions of the protocol, i.e., whether the system should lead to
a solved child system.
This process takes expert insight into the protocol, as the user needs to correctly judge the constraint system's traces, and Tamarin's internals, as the user needs to understand why Tamarin fails to automatically prove these systems.

Once the user identifies where and why Tamarin fails to finish a proof, they can try
to work towards a successful proof by 1) changing the model in a way that makes
it easier
for Tamarin to verify its properties, while keeping the same scope of verification,
2) identifying a useful intermediate lemma that can be proven independently and
reduces the search space, or 3) discovering a proof strategy, i.e., a sequence of
constraint solving rule applications, that can be turned into a heuristic for
an automatic proof.

Specifying a heuristic can be done in numerous ways, ranging from fact-specific
annotations in the model that (de)prioritize applying constraint solving rules to
these facts to providing a so-called
\emph{oracle}, which is an executable script, often written in Python,
called by Tamarin whenever the constraint solving rules need to be ranked.
However, \emph{tactics}~\cite{racou2024} have recently become
the standard way to provide heuristics to Tamarin. Tactics are a small
domain-specific language built into Tamarin that allows users to specify
a heuristic by ranking constraint solving rules based on which \emph{ranking
function} they match.
See~\Cref{fig:tamarin_tactic} for an example tactic from the 5G Handover XN model from~\cite{5Gformalanalysis21}.
A tactic has a name and multiple \emph{prio} or \emph{deprio} segments,
each containing functions that match the applicable
constraint solving rules. Rules are then ranked based on which
segment they match, and, should multiple rules match the same segment, based
on an optional, inbuilt \emph{presort} heuristic that ranks rules within the
same segment.

Most commonly, the ranking functions of tactics are regexes that match on the
pretty-printed string representation of the constraint solving rules, as
this gives users most control. One limitation of tactics is that they
are \emph{stateless} and only have access to the applicable constraint solving
rules of a constraint system and not the system itself, which could be useful for guiding
the proof search. As a result, tactics cannot specify that a rule should be
ranked higher if the constraint system contains a specific fact or if it has already been applied multiple times.

\begin{figure}
    \input{./img/tactic_content.tex}
\end{figure}

\subsection{A Reusable API for the Tamarin Prover}


In~\cite{5Ganalysis18}, the authors report the proof of the protocol's
properties ``require[d] dedicated and involved proof strategies ($\sim$1000 LoC
of Python)'' and that ``[t]he effort of writing such generic proof strategies
represents several person-months.''
Other Tamarin case studies do not specifically report on the time taken
for writing heuristics, but they do report on the overall time taken for
the proof effort. For instance, the authors of the WPA2 case
study~\cite{wpa2usenixsite} report 12 person-months of work,
\cite{cremers2023formal} reports 6-7 person-months of work, and
\cite{cremers2025breaking} reports 3-4 person-months
of work building on the models of~\cite{cremers2023formal}.
From corresponding with the authors, we know that a significant portion of
the reported time was spent on manually searching for candidate heuristics,
making it clear that this is a significant bottleneck for Tamarin verification.

The cause of this bottleneck is the lack of search \emph{over} the
applicable constraint solving rules: If Tamarin's greedy search fails to
find a proof with the inbuilt heuristics, users have to manually search
over the applicable rules to identify promising proof strategies.

While we could implement a custom proof search over the applicable rules by directly modifying Tamarin's source code, we have higher aspirations:
We want to empower Tamarin's users to implement their own proof search algorithms, rather than being limited to the one we provide.

To this end, we implement a reusable API for Tamarin that allows users to interact
with Tamarin's proof search without modifying its source code,
significantly lowering the entry barrier for implementing new proof search algorithms.
Our API turns Tamarin into a stateless REST server that exposes endpoints for
querying the initial constraint system and its applicable rules, for 
executing proof methods, and for validating the final proof.
This way, users can implement their own proof search algorithms by simply
implementing a client that queries Tamarin's REST API, without having to
interact with the internals of Tamarin or modifying its source code.

As a side effect, this also allows users to implement their proof search
algorithms in any programming language they want instead of being constrained to
Haskell by Tamarin, and
to implement heuristics that are more complex than what Tamarin's tactics
allow, for instance, by inspecting the current constraint system or the history
of applied rules in addition to the applicable rules.

Our API consists of three main endpoints: 1) \z{getInitialSystem},
which returns the initial constraint system of the proof search and its
applicable rules, 2) \z{executeMethod}, which takes as input a constraint
system and a constraint solving rule to apply to it, checks whether the rule
is applicable, and then returns the resulting
child constraint systems and their applicable rules, and 3) \z{checkProof},
which takes as input a proof, checks whether it is a valid proof of the desired
property, and returns a Tamarin \z{spthy} file containing the proof as well as
metadata about the proof such as the size.

Our API can now be used to implement a proof search as follows: The client
starts by querying the \z{getInitialSystem} endpoint to get the initial
constraint system of a proof and its applicable rules. It then selects a rule
to apply, for instance, by using a heuristic or by searching over the applicable
rules, and queries the \z{executeMethod} endpoint to apply this rule and get
the resulting child constraint systems and their applicable rules. This process
is repeated until a solved constraint system is found or the absence of a
counterexample is confirmed, at which point the client can construct a proof tree
and query the \z{checkProof} endpoint to check whether it is a valid proof.
\Cref{fig:tamarin_api} visualizes this interaction between the client and
Tamarin. Notably, the search tree of the proof search is now stored by the client instead of Tamarin, allowing the client to implement any proof search algorithm. 

\begin{figure}
    \input{./img/api_content.tex}
\end{figure}


\section{Reinforcement Learning for Tamarin}
\label{sec:ml}
Tamarin's built-in proof search greedily selects the top-ranked applicable proof method at each step.
This approach is efficient when the heuristic is informative, but it often leads to suboptimal and non-terminating search when the heuristic is imperfect.
In contrast, our approach formulates proof search as a full graph exploration problem where not only AND nodes (case splits) but also OR nodes (choosing which constraint to solve) are part of the search space.
The selection of which proof method to apply is guided by a learned policy and value function, optimized via Monte Carlo Tree Search (MCTS), building on the AlphaZero paradigm of combining neural network guidance with tree search~\cite{silverGeneralReinforcementLearning2018}, recently applied to formal mathematics by AlphaProof~\cite{hubertOlympiadlevelFormalMathematical2026}.
The reinforcement learning framework maps naturally onto Tamarin's proof search: States correspond to constraint systems, actions correspond to the applicable proof methods at a given proof state, and rewards encode proof progress signals.
Our system benefits from a self-improving loop: Parallel workers run MCTS searches on different lemmas using the API from~\Cref{sec:tamarin_env} and collect training examples from successful subproofs. These examples then update the neural network, which in turn guides the current and subsequent searches more effectively.

\subsection{Proof Search as Graph Exploration}
\label{subsec:ml_mcts}

\begin{figure}[t]
  \centering
  \includegraphics[width=\columnwidth]{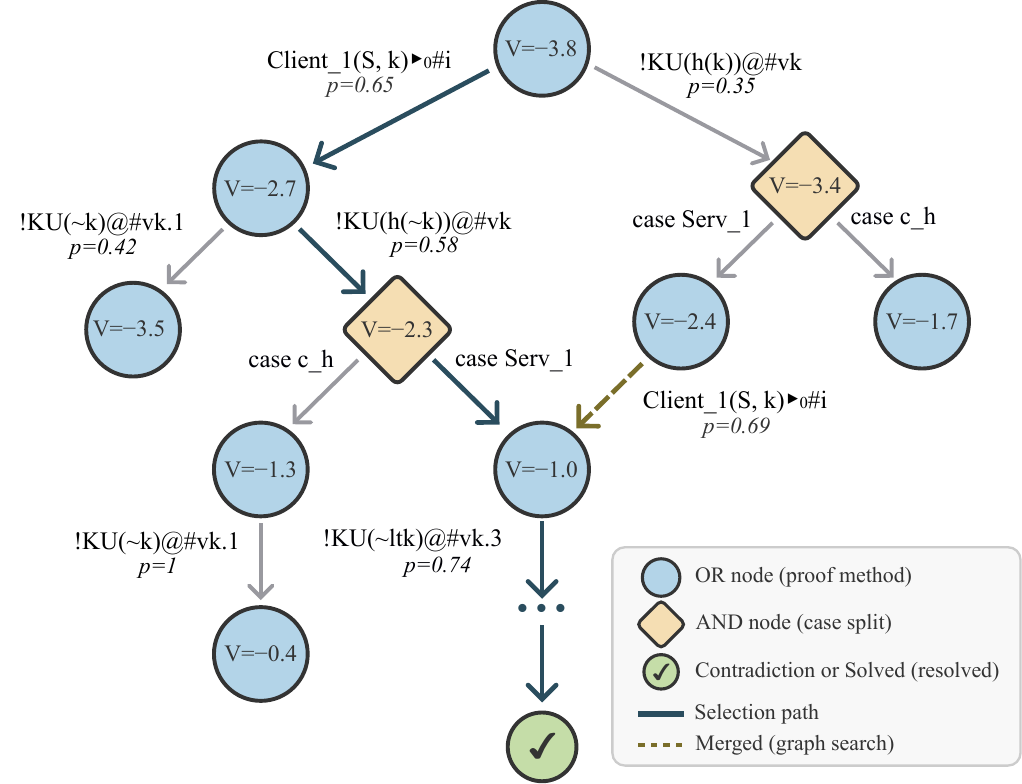}
  \Description{An AND/OR search graph with five levels, rooted at a blue circular OR node. The root has two outgoing edges, each labeled with a proof method name and a neural network prior probability. The left edge leads to another OR node, which itself branches into a leaf OR node on the left and a yellow diamond-shaped AND node on the right. The right edge from the root leads to a second AND node, which splits into two OR nodes via case Serv\_1 and case c\_h. Each AND node enforces a case split, requiring all its children to be resolved. The deeper AND node similarly splits into two OR nodes; one of these continues downward through further proof method edges to a leaf OR node, while the other is followed by an ellipsis leading to a terminal green checkmark node, indicating a resolved proof goal. All OR nodes display a backed-up value estimate V, and all edges show the proof method and its prior p. A thick dark path traces the PUCT selection trajectory from the root down to the resolved checkmark. A dotted edge connects an OR node in the right subtree to an OR node in the left subtree, illustrating graph-search deduplication where two different proof paths reach the same constraint system.}
  \caption{AND/OR search graph for the \emph{Tutorial} protocol model, lemma \emph{Client Session Key Secrecy} after several MCTS iterations. OR nodes (circles) represent constraint systems where the agent selects a proof method; AND nodes (diamonds) arise from case splits requiring all children to be resolved. Edges carry the proof method and neural network prior~$p$. Node values $V$ reflect backed-up estimates. The thick path shows PUCT selection; the dotted edge illustrates graph-search deduplication.}
  \label{fig:mcts-graph}
\end{figure}

We adapt MCTS to Tamarin's proof search by modeling the search space as a graph with two types of nodes:
\emph{OR nodes} represent constraint systems where the agent must choose which proof method to apply next. Any single choice that leads to a proof suffices.
\emph{AND nodes} arise from case splits when the application of a proof method produces multiple child constraint systems.
Their semantics change depending on whether there exists a solution to the constraint system or not: If there is a solution, for instance because a counterexample to a universal property is found, then the case split acts as a disjunction, as it suffices to find a solution in any of the branches. If there is no solution because the constraint system is ultimately contradictory, then the case split acts as a conjunction, as all branches must be proven contradictory to show the absence of a solution.
Because it is not known a priori whether a property has a solution or not, we conservatively treat all case splits as AND nodes.
Once a (sub-)proof is found, the search can identify which branches of the AND nodes are relevant and update their status accordingly.
\Cref{alg:mcts} gives an overview of the search procedure for a single lemma and \Cref{fig:mcts-graph} illustrates the resulting search graph.

\begin{figure}[t]
\input{img/pseudocode.tex}
\caption{Main MCTS loop. The four phases (Selection, Expansion, Backup, and Extraction) are explained in detail in Section~\ref{subsec:ml_mcts}.}\label{alg:mcts}
\end{figure}

The following paragraphs detail each phase.

\paragraph{Selection.}
We traverse the graph from the root. At each OR node we select the child that maximizes a
PUCT-style score:
\[
  \text{PUCT}_{\text{OR}}(s,a) \;=\;
    \underbrace{\gamma^{\,-1 - V(s,a)}}_{\text{value score}}
    \;+\;
    \underbrace{c(s)\;\cdot\;p(a \mid s)}_{\text{prior score}},
\]
where $V(s,a)$ is the aggregated value of the child node reached by taking action~$a$ in state~$s$, $\gamma \in (0,1]$ is a value discount that
controls sensitivity to value differences, $p(a \mid s)$ is the prior probability assigned by the neural network to action~$a$ in state~$s$, and
\[
  c(s) \;=\; \left[\log\!\left(\frac{\sum_b N(s,b) + c_{\text{base}} + 1}{c_{\text{base}}}\right) + c_{\text{init}}\right]
  \cdot \frac{\sqrt{\sum_b N(s,b)}}{N(s,a) + 1}
\]
is the exploration coefficient, with $N(s,a)$ the number of times action~$a$ has been selected from state~$s$ and $\sum_b N(s,b)$ the total number of actions taken from~$s$, summed over all available actions~$b$.
For unvisited actions, the value is estimated pessimistically as $V(s) - \delta$, where
$\delta$ is a configurable penalty.
Following AlphaProof~\cite{hubertOlympiadlevelFormalMathematical2026}, we replace the linear value term of the AlphaZero PUCT formula with an
exponential discount to handle the unbounded returns in proof search and maintain a consistent scale between the value and prior terms.
The discount $\gamma$ additionally controls sensitivity: $\gamma$ close to $1$ compresses differences (prior-dominated), while smaller values amplify them (value-dominated).

At AND nodes, following AlphaProof~\cite{hubertOlympiadlevelFormalMathematical2026}, priors are uniform, and the exploration coefficient is scaled by $c_{\text{and}}$ to control exploration at AND nodes separately from OR nodes:
\[
  \text{PUCT}_{\text{AND}}(s,a) \;=\;
    \gamma^{\,-1 - V(s,a)}
    \;+\;
    c_{\text{and}} \cdot c(s) \cdot \frac{1}{|A(s)|}
\]
Contrary to AlphaProof, we do not invert the value score at AND nodes for two reasons: First, AlphaProof's prioritization of the hardest sub-goal stalls finding subproofs from which we can learn (See \textit{Extraction}). Second, it hurts existential lemmas, where case splits are disjunctions and it is more efficient to find any single subproof first.

\paragraph{Expansion \& Evaluation.}
When selection reaches an unexpanded OR node, we first query the neural network to obtain a prior
distribution $\mathbf{p}$ and a value estimate~$v$ for the state.
We then apply the top N applicable proof methods to Tamarin, obtaining the child
constraint systems for each method. This is different from both AlphaZero, which expands only a single child per node, and AlphaProof, which expands all children.
We decided on this middle ground to eagerly detect contradictions and solved states and to generate training data while keeping the compute overhead manageable with parallelization.
Each proof method application yields a set of children.
More than one child indicates a case split, which we handle by creating an AND node with uniform prior and adding all child constraint systems as OR nodes.
A proof method that produces zero children is contradictory by design, and a single child system may be either a terminal state (solved or contradictory) or an OR node.
Terminal nodes receive a value of $1.0$ instead of a neural estimate.
Proof methods that exceed a per-call timeout are retried with a longer timeout, and, if they still fail, they are excluded from the search.

\paragraph{Backup.}
After expansion, values and status flags (contradictory, solved) are propagated back along the
traversed path.
OR nodes aggregate child values as a weighted running mean:
$$
  V_{\text{OR}} \;=\; \frac{v + \sum_{a \in \text{visited}} N(s,a) \cdot (R(s,a) + V(s,a))}
                           {1 + \sum_{a} N(s,a)},
$$
where $v$ is the network's initial value estimate for the node, $N(s,a)$ is the number of
times action~$a$ has been selected from state~$s$, $R(s,a)$ is the edge reward (\Cref{subsec:ml_reward}), and $V(s,a)$ is the
aggregated value of the child reached via action~$a$.
An OR node becomes contradictory if any child is contradictory, and it becomes solved if any child is solved.

AND nodes take the minimum value over visited, non-contradictory children,
reflecting that the overall difficulty is determined by the hardest remaining sub-case.
Unvisited children are treated optimistically (assigned the maximum value~$1$), so the AND node's value reflects only the known hardest subgoal.
An AND node becomes contradictory only when \emph{all} its children are contradictory, but it becomes solved if any child is solved.

\paragraph{Extraction.}
We extract training examples from the search whenever a node becomes resolved (contradictory or solved), rather than waiting until the search terminates.
Extracting partial subproofs mid-search, rather than waiting for a complete proof, provides training signal early and allows the network to improve while the search is still running.
To avoid contamination of the training data with the network's own value estimates, we recompute the objective value targets $v_t(s)$ for all OR nodes by backing up values analogously to the backup phase, but without averaging with the network's own value estimates.
Training examples are tuples $(s, a, v_t)$, where $s$ is the proof state, $a$ is the index of the selected action, and $v_t$ is the value target.

When the root node is finally resolved, the extracted proof tree is validated independently by sending it to Tamarin's \emph{checkProof} endpoint via the API, and the search is finished.

\paragraph{Graph Search.}

Contrary to the standard MCTS formulation and AlphaZero~\cite{silverGeneralReinforcementLearning2018}, we perform a Monte Carlo graph search~\cite{wu_graphsearch,czechMonteCarloGraphSearch2020}, since the same constraint system can be reached via different proof paths; for instance, through different interleavings of the
same proof methods.
While this turns the search tree into a directed acyclic graph (DAG), graph search does not fundamentally change the MCTS algorithm. Its main advantage is efficiency: By merging duplicate states, we significantly reduce the search space by avoiding redundant exploration.
We find in our experiments that usually around $2-5\%$ of the nodes in the search are shared, with some protocols up to $25\%$.

\subsection{Neural Network Architecture}
\label{subsec:ml_architecture}

The Reinforcement Learning agent relies on a neural network to guide its search by providing a policy and a value estimate for each proof state, taking an observation of the state as input.

\paragraph{State Encoding.}
We experimented with different full-state encodings, including a Graph Neural Network, but we quickly encountered scalability issues due to the size of the constraint system.
As a consequence, we instead abstract from the full state, using the set of applicable proof methods as the observations.
This aligns with how Tamarin's heuristics work.

The observable proof state is encoded by a Transformer-based neural network.
Each proof method is represented as a sequence of its method type and arguments, similar to its representation in the Web interface, and tokenized using a standard subword tokenizer~\cite{liuRoBERTaRobustlyOptimized2019}.
Token sequences are embedded via a learned embedding layer into $d$-dimensional vectors $\boldsymbol{x}_t \in \mathbb{R}^d$, augmented with sinusoidal positional encodings, then processed by a standard Transformer encoder.
The resulting token-level representations are mean-pooled over each proof method's tokens, yielding a single vector $\boldsymbol{e}_i \in \mathbb{R}^d$ per proof method.
\paragraph{Policy Head.}
For a state $s$ consisting of $L$ proof methods, the policy head scores each of them against the full set, using a pointer attention mechanism~\cite{vinyalsPointerNetworks2015}:
Each method's encoding is projected into a key $\boldsymbol{k}_i \in \mathbb{R}^d$.
A shared query $\boldsymbol{q} \in \mathbb{R}^d$ summarizing the entire state is obtained by summing all method encodings (after a nonlinear transformation) and applying a linear projection on the result.
The unnormalized score (i.e., logit) for method~$i$ is
$
  \ell_i = \boldsymbol{w}^\top \tanh(\boldsymbol{k}_i + \boldsymbol{q}) \in \mathbb{R},
$
with $\boldsymbol{w} \in \mathbb{R}^d$. We then apply a log-softmax on $\{\ell_i\}_{i=0}^{L-1}$ to normalize the logits into a prior probability distribution $p(\cdot \mid s) \in \mathbb{R}^L$ over proof methods.


\paragraph{Value Head.}
The value head is a separate projection layer $V\!:\!~\mathcal{S}\times \mathcal{A} \to \mathbb{R}$, where $\mathcal{S}$ and $\mathcal{A}$ are the state space and action space, respectively. The $V$ layer is equipped with the same pointer-network architecture as the policy head to produce per-method scores. 
The scores are then mean-pooled across all methods in the state to obtain a single scalar value estimate.

Both heads share the same Transformer encoder, forming an actor-critic architecture on shared
representations.

\subsection{Integrating Heuristic Knowledge}
\label{subsec:ml_heuristic}

To leverage existing domain knowledge, we combine the learned prior with a rank-based bias derived from Tamarin's heuristic ordering.
We exploit this ordering by adding a rank-dependent bias to the network's logits before
normalization:
\[
  \mathbf{p}_{\text{combined}}
  \;=\; \operatorname{softmax}\!\left(\frac{\ell - \lambda \cdot r}{T}\right),
\]
where $\ell$ are the raw model logits, $r = (0, 1, 2, \ldots)$ are the
rank positions from the heuristic ordering, $\lambda$ controls the heuristic's influence, and $T$ is
a temperature parameter that flattens the combined distribution, making it more uniform.
Setting $\lambda = 0$ yields a purely learned prior; increasing $\lambda$ interpolates toward the
heuristic ordering.
If the inbuilt heuristic is informative, this bias can help the search converge faster by guiding it toward promising proof methods early on.
Vice versa, if the heuristic is misleading, the heuristic can either be switched off manually or the learned prior can override it.
A human-engineered heuristic can be integrated in the same way, allowing the framework to benefit from human insights along the process.

\subsection{Reward Shaping}
\label{subsec:ml_reward}

Each edge in the search graph carries a reward that the agent seeks to maximize.
Following AlphaProof's approach of penalizing each proof step, we use a base cost of $-1$ per step.
By encouraging shorter proofs, we implicitly encourage termination in general, as diverging paths or loops accumulate large negative rewards.
We augment this base cost with three domain-specific penalty terms, each normalized to a comparable range, yielding the overall reward:
\[
  R(s,a) \;=\; -(1 + \beta \cdot b + \alpha \cdot t + \tau \cdot h),
\]
where $b$, $t$, and $h$ are the branching, soft-time, and hard-timeout penalties defined below, and $\beta$, $\alpha$, $\tau$ control their respective influence.
Terminal edges (contradictions, solved states) and edges to AND nodes carry only the base cost of~$-1$.

\paragraph{Branching Penalty.}
The branching penalty~$b$ penalizes transitions that increase the number of applicable proof methods relative to the worst-case branching factor on the current path, steering the search toward states with fewer applicable methods, which tend to be closer to a terminal state.
\paragraph{Soft-Time Penalty.}
The soft-time penalty captures the wall-clock time~$\Delta t$ Tamarin takes to expand a node, clipped and normalized: $s = \min(\Delta t,\, t_{\text{clip}}) / t_{\text{clip}}$.
This guides the search away from proof methods that are computationally expensive to execute.
\paragraph{Hard-Timeout Penalty.}
If a proof method exceeds the per-call timeout, we retry with a longer sequential timeout to avoid permanently excluding reachable states.
Methods that eventually succeed under the extended timeout receive a fixed penalty~$h$, discouraging paths that repeatedly trigger expensive Tamarin calls.
If the extended timeout also fails, the proof method is permanently excluded.

\subsection{Training}
\label{subsec:ml_training}

\begin{figure}[t]
  \centering
  \includegraphics[width=\columnwidth]{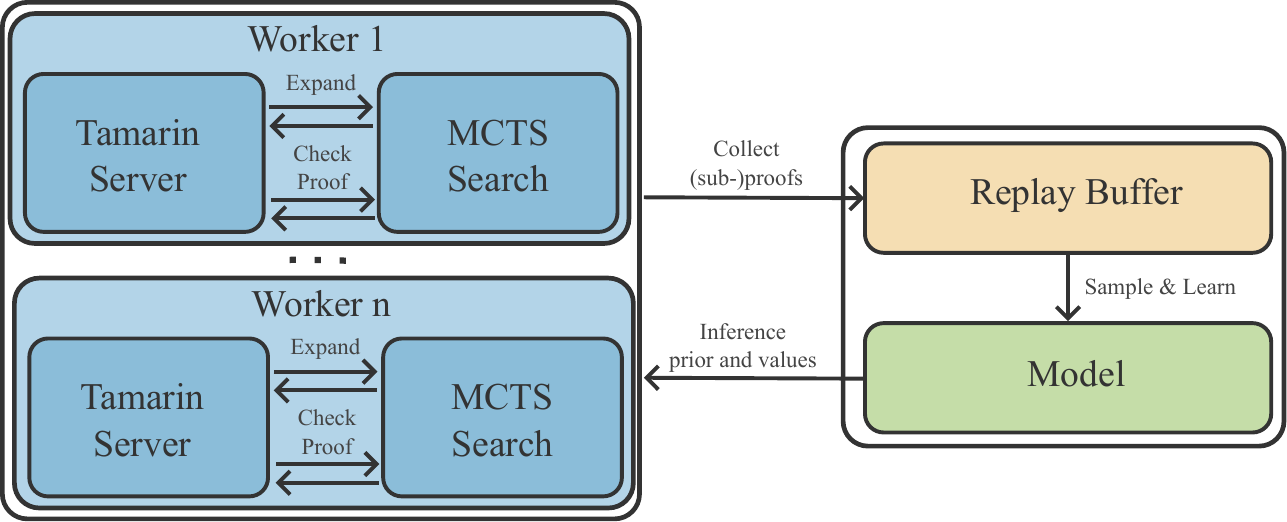}
  \Description{Block diagram of the training architecture. On the left, multiple workers (Worker 1 through Worker n) are shown, each containing a Tamarin Server and an MCTS Search component connected by bidirectional Expand and Check Proof arrows. On the right, the ML Model side contains a Replay Buffer and a Model connected by a Learn arrow. Between the two sides, a Collect sub-proofs arrow flows from the workers to the Replay Buffer, and an Inference prior and values arrow flows from the Model back to the workers.}
  \caption{Training architecture overview. Multiple workers run MCTS searches in parallel, each communicating with its own Tamarin server instance. Workers collect training examples from subproofs and send them to a shared replay buffer. The model learns from buffer samples and provides inference (prior and value estimates) back to the workers.}
  \label{fig:training-overview}
\end{figure}

Training follows an online self-play loop (\Cref{fig:training-overview}): Parallel workers run independent MCTS searches, extract training examples from successful subproofs (\Cref{subsec:ml_mcts}), and feed them into a shared replay buffer. The trainer process continuously samples from this buffer and updates the neural network, whose improved predictions immediately guide ongoing and subsequent searches.
Unlike AlphaProof, which only learns from fully solved problems, our online loop generates training signal from partial subproofs, making training feasible without extensive pretraining.

\paragraph{Replay Buffer.}
Training examples are stored in a FIFO buffer. We sample from the buffer with a probability inversely proportional to the number of times an example has been drawn, to mitigate staleness and improve data efficiency. We sample from this buffer continuously under two conditions: (1)~the least seen example has been seen fewer than $M$ times, and (2)~the buffer has been filled to a fraction $f$ of its capacity, to ensure a minimum level of diversity in the training data. Both $M$ and $f$ are configurable parameters.

\paragraph{Losses.}
The training objective combines a cross-entropy loss on the policy head (encouraging the network to
predict the MCTS-selected action) with a mean squared error loss on the value head (encouraging the
network to predict the backed-up value target).
We use the Adam optimizer with gradients clipped at a global norm of $1.0$.

\paragraph{Search Budget.}
Each MCTS search is allocated a budget of $B$ node expansions.
If the proof is not found within the budget, the search terminates and a new search begins with an
increased budget $B' = \lfloor B \cdot \phi \rfloor$, where $\phi > 1$ is a configurable growth
factor.
This allows the system to attempt quick proofs first and progressively invest more computation in
harder lemmas.

\paragraph{Multi-worker Parallelism.}
Multiple worker processes run MCTS searches in parallel, while neural network inference and gradient updates are
centralized in the trainer process on a single GPU.
Workers send proof states to the trainer via pipes and receive policy and value predictions in
return, always using the latest model weights without explicit synchronization.
Collected training examples flow back through a shared queue into the replay buffer.
This architecture allows us to scale the number of workers independently of the GPU resources.

\paragraph{Multi-lemma Training.}
Workers can operate on different lemmas or even different protocols while sharing a single neural network, enabling
transfer of learned proof strategies across related properties.
The trainer dynamically schedules workers across lemmas, starting new searches as previous ones
complete or exhaust their search budget. The scheduler selects lemmas in a round-robin fashion until a number of searches or proofs are completed, so easier lemmas are solved first and contribute more training examples early on, while harder lemmas receive more attention as the training progresses and the network improves.

\section{Case Studies and Experimental Results}
\label{sec:case_studies}
To demonstrate the efficacy of our approach, we picked 16 case studies from Tamarin's main repository and recent publications. 
These case studies range from classical protocol models that Tamarin quickly solves by default, over models where some human guidance was necessary to guide the proof search, to state-of-the-art models from recent publications, where person-months of effort went into finding heuristics~\cite{wpa2usenixsite,cremers2023formal,cremers2025breaking,5Ganalysis18}.

We find that our approach solves more lemmas automatically than Tamarin's default heuristics and that it produces shorter proofs than both the default heuristics and the original human-written heuristics.
Moreover, we show that our approach generalizes across different lemmas within the same protocol, using knowledge gained from previous proof attempts to assist in proving future lemmas.


In the following, we first provide an overview of the case studies we selected and then describe our experiments and findings.

\paragraph{Case Studies.}
We group our case studies into three categories: The first category consists of classical security protocol models from Tamarin's main
repository~\cite{tamarinexamples}, which are quickly solved by Tamarin's default heuristics and require no human guidance.
The second category consists of more complex models, such as WireGuard~\cite{donenfeld2017formal}, the Signal model from~\cite{linker2025looping}, and 5G Handover EPS N26~\cite{5Gformalanalysis21}, for which some human effort was necessary to guide the proof search.
The third category consists of state-of-the-art models like 5G AKA~\cite{5Ganalysis18}, SPDM~\cite{cremers2023formal,cremers2025breaking}, and WPA2~\cite{wpa2analysis}, where person-months of effort went into finding heuristics~\cite{wpa2usenixsite,cremers2023formal,cremers2025breaking, 5Ganalysis18}.

\paragraph{Experimental Setup.}
Our experiments are structured as follows:
We first establish Tamarin's baseline performance on our case studies by removing all human annotations and heuristics that influence the search heuristics, but we keep helper lemmas and the corresponding \emph{reuse} and \emph{hide\_lemmas} annotations.
While identifying key helping lemmas is also an important aspect of the proof effort, we consider it complementary to our work and only focus on the proof search.
We then run Tamarin on these models using its default search strategy (IDDFS) with the default heuristic \z{s} and the heuristic \z{c} with a timeout of two hours, respectively four hours for especially large protocols (SPDM and WPA2).
We also chose to include the \z{c} heuristic in our \emph{baseline} performance because it ranks applicable proof methods based on how long they were applicable to the constraint system and, as a result, never indefinitely delays applying one, making it the best attempt at searching over the applicable proof methods that Tamarin currently has. 

We also run Tamarin on the original, unaltered case studies in an attempt to reproduce the results reported in the respective papers on our Tamarin version.
These runs reflect the \emph{original} performance of Tamarin on these case studies, usually with significant human engineering effort required to create effective heuristics.
For the classical models, the original performance is often achieved without human heuristics, as they were not necessary to solve all lemmas in the case study. For some case studies, we fail to reproduce the reported results from the original papers (See~\Cref{app:original_results} in the Appendix for details).

We then train and evaluate our approach as described in~\Cref{subsec:ml_training}.
We experiment with multiple configurations, varying key hyperparameters such as the heuristic influence $\lambda$, exploration coefficients, and search budget $B$. We report the details in~\Cref{tab:hyperparameters} in the Appendix.
We report the results of the best single configuration.

We use an Intel(R) Xeon(R) CPU E5-4650L 2.60GHz machine with 1TB of RAM and 8 cores per Tamarin call for the Tamarin baseline experiments.
For our reinforcement learning setting, we use one NVIDIA A100 GPU with 40GB of memory, up to 64 cores AMD Rome 7742, 2.25 GHz, and 250GB of RAM.
While we have not systematically evaluated on consumer-grade CPU-only hardware, running without a GPU is generally possible. We recommend at least 60 GB of memory.
Due to reduced worker parallelization and slower training, longer run times are to be expected.

\begin{table}[htbp]
\centering
\caption{Overview of how many lemmas our RL system (best single configuration), and Tamarin's \z{c} and \z{s} heuristics, solve in each case study.
The Total column shows the number of total lemmas in this category, with the \emph{original} heuristics in parentheses when fewer than the total.
See the paragraph \textbf{Less Effort} in~\Cref{par:outperforming-baseline} for a discussion of this table.}
\label{tab:lemma-completion}
\begin{tabular}{@{}l c c c c c@{}}
\toprule
Protocol & Type & Total & \z{c} & \z{s} & RL \\
\midrule
\multicolumn{6}{c}{\textbf{Classical Models}} \\
\midrule
\multirow{2}{*}{FOO Eligibility} & $\forall$ & 2 & 2 & 2 & \textbf{2} \\
 & $\exists$ & 1 & 1 & 1 & \textbf{1} \\
\lightrule
KAS2 eCK & $\forall$ & 1 & 1 & 1 & \textbf{1} \\
\lightrule
NAXOS eCK & $\forall$ & 1 & 0 & 1 & \textbf{1} \\
\lightrule
NAXOS eCK PFS & $\forall$ & 1 & 1 & 1 & \textbf{1} \\
\lightrule
\multirow{2}{*}{Tutorial} & $\forall$ & 3 & 3 & 3 & \textbf{3} \\
 & $\exists$ & 1 & 1 & 1 & \textbf{1} \\
\lightrule
UM PFS & $\forall$ & 2 & 2 & 2 & \textbf{2} \\
\lightrule
\multirow{2}{*}{YubiKey} & $\forall$ & 3 (0) & 0 & 0 & \textbf{1} \\
 & $\exists$ & 1 & 1 & 1 & \textbf{1} \\
\lightrule
YubiKey HSM & $\forall$ & 4 & 4 & 4 & \textbf{4} \\
\midrule
\multicolumn{6}{c}{\textbf{Complex Models}} \\
\midrule
\multirow{2}{*}{5G Handover EPS N26} & $\forall$ & 7 & 2 & 2 & \textbf{7} \\
 & $\exists$ & 1 & 1 & 1 & \textbf{1} \\
\lightrule
PKCS11 AEAD & $\forall$ & 4 & 3 & 4 & \textbf{4} \\
\lightrule
\multirow{2}{*}{Signal} & $\forall$ & 4 (1) & 1 & 1 & \textbf{4} \\
 & $\exists$ & 1 (0) & 1 & 0 & \textbf{1} \\
\lightrule
\multirow{2}{*}{Wireguard} & $\forall$ & 6 & 4 & 6 & \textbf{6} \\
 & $\exists$ & 2 & 1 & 2 & \textbf{2} \\
\midrule
\multicolumn{6}{c}{\textbf{State-of-the-Art Models}} \\
\midrule
\multirow{2}{*}{5G AKA} & $\forall$ & 9 (8) & 5 & 6 & \textbf{9} \\
 & $\exists$ & 4 & 3 & 3 & \textbf{4} \\
\lightrule
\multirow{2}{*}{5G Handover XN} & $\forall$ & 13 & 3 & 3 & \textbf{8} \\
 & $\exists$ & 7 & 0 & 0 & \textbf{1} \\
\lightrule
\multirow{2}{*}{SPDM} & $\forall$ & 31 & 24 & 28 & \textbf{30} \\
 & $\exists$ & 18 (17) & 14 & 15 & 7 \\
\lightrule
\multirow{2}{*}{WPA2} & $\forall$ & 73 & 41 & 32 & \textbf{71} \\
 & $\exists$ & 12 (2) & 2 & 2 & 2 \\
\bottomrule
\end{tabular}
\end{table}

\paragraph{Less Effort.}
\label{par:outperforming-baseline}

In \Cref{tab:lemma-completion} we summarize the number of lemmas solved by each approach.
The \emph{Total} column shows the total number of lemmas in each case study, split into universal and existential lemmas.
Our approach consistently matches or outperforms Tamarin's baseline heuristics, with the advantage becoming more pronounced on the complex and state-of-the-art models.
On the classical models, we match the baseline and solve one additional lemma on YubiKey\footnote{This is \emph{not} the YubiKey model verified in the SmartVerif paper~\cite{xiongSmartVerifPushLimit2020} (see~\Cref{sec:related_work}), as their
version appears to be a SAPIC+~\cite{cheval2022sapic+} model that was translated to Tamarin.}.

On the complex models, the gap widens substantially: Our approach solves all lemmas on 5G Handover and Signal, whereas the baselines solve only a fraction, that is 3 out of 8 on 5G Handover, and on Signal 1 out of 5 even with the original heuristic, while 2 out of 5 with the \z{c} heuristic. On WireGuard and PKCS11, we match the baseline with all lemmas solved.

On the state-of-the-art models, we continue to outperform the baseline significantly, solving 73 of 85 WPA2 lemmas (baseline: 43) and 9 of 20 5G Handover XN lemmas (baseline: 3). On the 5G AKA model, we solve all 13 lemmas, whereas the baseline solves only 9, and the original heuristics 12.
On WPA2 we solve an additional 30 lemmas compared to the best baseline heuristic and almost match the original heuristics (us 73 vs. original 75), without needing any person-months of manual effort to design heuristics. On top of the results in~\Cref{tab:lemma-completion} (best single configuration), a second RL configuration also solves one additional WPA2 lemma, bringing the union to 74.

On the SPDM model, we do not fully match the baseline heuristics, solving 37 out of 49 lemmas vs. 43.
We attribute this to the size and complexity of the model: Single Tamarin calls can take up to 10 minutes (timeout), and sequential backup calls can add up to 90 minutes per node expansion.
This severely limits exploration, and a greedy selection strategy, as the baseline heuristics employ, can perform better in such a setting. In addition, the SPDM model contains a large number of existential lemmas, which we find particularly difficult for our approach, as described in the Limitations.

In summary, our approach matches or significantly outperforms the default heuristics across all models except the SPDM model, proving statements that previously required human guidance.
Our approach nearly matches the original, human-engineered heuristics on the most complex models, without any manual engineering effort for heuristics.
Moreover, our approach outperformed the original heuristics on several models where the originally reported performance was no longer reproducible with the current Tamarin version, highlighting the brittleness of changes in Tamarin versions and the advantage of our approach being more robust to such changes.
While our approach does not yet fully automate proof search for the most complex models, it substantially reduces the human effort required.

\begin{table}[htbp]
  \centering
  \caption{Average proof size for the lemmas solved by both the respective baseline
  and our approach (RL). Best configuration per lemma.
  See the paragraph \textbf{Shorter Proofs} in~\Cref{par:shorter-proofs} for
  a detailed explanation and discussion of this table.}
  \label{tab:proofsize-overview}
  \begin{tabular}{@{}lrrrr@{}}
    \toprule
    & \multicolumn{2}{c}{\z{c}/\z{s} $\cap$ RL} & \multicolumn{2}{c}{Orig.\ $\cap$ RL} \\
    \cmidrule(lr){2-3}\cmidrule(lr){4-5}
    Protocol & \multicolumn{1}{c}{\z{c}/\z{s}} & \multicolumn{1}{c}{RL} & \multicolumn{1}{c}{Orig.} & \multicolumn{1}{c}{RL} \\
    \midrule
    \multicolumn{5}{c}{\textbf{Classical Models}} \\
    \midrule
    FOO Eligibility & 21.0 & \textbf{20.3} & 21.0 & \textbf{20.3} \\
    KAS2 eCK & 12.0 & \textbf{11.0} & 16.0 & \textbf{11.0} \\
    NAXOS eCK & 134.0 & \textbf{117.0} & 134.0 & \textbf{117.0} \\
    NAXOS eCK PFS & 13.0 & \textbf{8.0} & 13.0 & \textbf{8.0} \\
    Tutorial & 9.0 & \textbf{8.2} & 9.0 & \textbf{8.2} \\
    UM PFS & 8.0 & 8.0 & 8.0 & 8.0 \\
    YubiKey & 12.0 & 12.0 & 12.0 & 12.0 \\
    YubiKey HSM & 378.8 & \textbf{20.8} & 570.0 & \textbf{20.8} \\
    \midrule
    \multicolumn{5}{c}{\textbf{Complex Models}} \\
    \midrule
    5G Handover EPS N26 & 9531.7 & \textbf{17.3} & 85.8 & \textbf{24.4} \\
    PKCS11 AEAD & 471.8 & \textbf{245.2} & 471.8 & \textbf{245.2} \\
    Signal & 18.0 & \textbf{15.5} & 20.0 & 20.0 \\
    Wireguard & 77.4 & \textbf{49.1} & \textbf{47.8} & 49.1 \\
    \midrule
    \multicolumn{5}{c}{\textbf{State-of-the-Art Models}} \\
    \midrule
    5G AKA & 1679.0 & \textbf{148.8} & \textbf{73.3} & 118.6 \\
    5G Handover XN & 26.3 & \textbf{20.7} & 102.8 & \textbf{48.3} \\
    SPDM & 94.9 & \textbf{64.6} & 106.1 & \textbf{70.8} \\
    WPA2 & 903.7 & \textbf{34.8} & 81.9 & \textbf{52.0} \\
    \bottomrule
  \end{tabular}
\end{table}

\paragraph{Shorter Proofs.}
\label{par:shorter-proofs}
While proof size is not the primary optimization objective when verifying security protocols, shorter proofs are generally preferable as they are easier to understand and easier to maintain.
Similarly, counterexamples are easier to interpret when they are smaller, making it easier to identify and fix bugs in both the model and the real-world protocol.
Our approach finds significantly shorter proofs than the baseline heuristics and often shorter proofs than the original, human-engineered heuristics.
The columns in \Cref{tab:proofsize-overview} show the average proof size for the lemmas solved by the respective approach.
For a fair comparison, the columns \emph{\z{c}/\z{s} $\cap$ RL} show the average proof size for those lemmas that were solved by both our RL approach and either respective baseline heuristics, whereas the columns \emph{Orig.\ $\cap$ RL} show the averages for those lemmas that were solved by both our RL approach and Tamarin with the original heuristic.
Note that for some easier models, the original heuristics and the baseline heuristics are identical.

Compared to the baseline heuristics, we find shorter or equal sized proofs
across all models.
In some cases, the difference is substantial, with a factor of over 18 for YubiKey HSM and up to 500 for the 5G Handover model.
Such differences make shorter proofs not only a theoretical advantage but a practical necessity, as proofs of sizes in the thousands are impossible to interpret.

Our approach also often finds shorter proofs than the human-written original heuristics.
Most notably, we produce shorter proofs on 5G Handover EPS N26 (factor 3.5), 5G Handover XN (factor 2.1), and WPA2 and SPDM (factor 1.5--1.6).
Thus, our proofs distill the key proof steps, making them easier to understand than the proofs produced by the baseline and original heuristics.

\paragraph{Execution Time.}
\label{par:training-time}
Since our approach interleaves training and search, the proof times include both neural network updates and proof search via Tamarin.
On the classical models, all lemmas are solved within a couple of minutes, and on the complex models, it takes between a few minutes (5G Handover EPS N26, WireGuard) and about two hours (Signal).
Notably, on 5G Handover EPS N26, our approach solves more lemmas faster than both of Tamarin's built-in heuristics: 3 minutes vs.\ 45 minutes (\z{c}) and 49 minutes (\z{s}).
On the state-of-the-art models, our search requires between 4.5 and 20 hours.
While this is a significant computational cost, it remains well within a practical time frame and automates a significant part of the person-months of manual heuristic engineering that were necessary to solve a similar number of lemmas in these models.
A full timing comparison is provided in~\Cref{app:additional_experimental_data}.


\paragraph{Limitations \& Future Work.}
\label{par:limitations}
When conducting our experiments, we observed that our approach underperforms on existential lemmas.
We believe that this is due to two factors: First, solutions to existential lemmas are usually much deeper in the search tree than proofs for universal lemmas. 
As the search space grows exponentially with the depth of the tree, search-based methods struggle to find deep solutions and thus to solve existential lemmas.
Consequently, search-based methods must be particularly well guided to find such deep solutions, which brings us to the second factor:
Our models learn from completed (sub-)proofs.
A subproof through finding a solved constraint system immediately finishes the proof on these lemmas, hence we cannot learn from this data within the same search.
Subproofs through contradictory constraint systems, on the other hand, occur rarely in existential lemmas, as the property is \emph{meant} to be satisfied, and thus the search generates relatively little training data for our model.
In turn, the search on existential lemmas is less informed, adding to the difficulty of finding solutions for them.
We consider specialized solutions for existential lemmas an interesting target for future work. Possible approaches are a different backup logic or a different paradigm for selecting training data, as well as a specialized reward signal for existential lemmas.

Another challenge we encountered is the amount of time it takes to execute proof steps with Tamarin.
For instance, for the SPDM model and 5G Handover XN model, we observed that a single node expansion can take up to 90 minutes, with an average of 30 minutes. This significantly limits the search space we can explore and thus the performance of our approach. This drawback affects any approach that searches through the proof methods.
Directing the search better would address this, e.g., through machine learning models with stronger initial knowledge from pretraining.
In this work, we have not yet established whether transfer learning across protocols, which is the basis for pretraining, is effective.
In contrast to related work in mathematical theorem proving, where the underlying deductive system is fixed and thus pretraining is possible, in Tamarin, each protocol model defines its own semantics and should be viewed as its own deductive system.
Thus, it is an open question whether transfer learning across protocols is effective.


\section{Related Work}
\label{sec:related_work}
\paragraph{Reinforcement Learning for Theorem Proving.}
In recent years, there has been increasing interest in using machine learning for mathematical reasoning and interactive theorem proving.
Approaches range from generating whole proofs~\cite{jiangDraftSketchProve2022,zhao2023decomposingenigmasubgoalbaseddemonstration,wang2023legoproverneuraltheoremproving} to iteratively selecting single proof steps~\cite{poluGenerativeLanguageModeling2020b,bansalHOListEnvironmentMachine2019a,lampleHyperTreeProofSearch2022a,gloeckleABELSampleEfficient2024a,hubertOlympiadlevelFormalMathematical2026}; our work falls in the latter category.
Much of this line of work traces back to the AlphaGo lineage~\cite{silverMasteringGameGo2017,silverMasteringChessShogi2017,silverGeneralReinforcementLearning2018}, which combines learned policy and value functions with MCTS.\
Notable examples include HTPS~\cite{lampleHyperTreeProofSearch2022a}, ABEL~\cite{gloeckleABELSampleEfficient2024a}, DeepSeek-Prover-V1.5~\cite{xinDeepSeekProverV15HarnessingProof2024}, and AlphaProof~\cite{hubertOlympiadlevelFormalMathematical2026}, adapting the AlphaZero framework to theorem proving by using an MCTS to search over applicable proof steps or tactics.
Similar to our work, the proof search is structured as an AND/OR tree: Proof states where tactics are generated and selected are OR-nodes, while case splits are AND-nodes.
The policy and value networks are trained on data generated by the MCTS itself, following the same self-play reinforcement learning paradigm as AlphaZero.
Different from our work, HTPS~\cite{lampleHyperTreeProofSearch2022a} and ABEL~\cite{gloeckleABELSampleEfficient2024a} extract training data only after a proof search ends, while AlphaProof~\cite{hubertOlympiadlevelFormalMathematical2026} additionally requires the root to be fully solved.
We extract training samples from sub-proofs mid-search to benefit from those learning signals as soon as possible.
Since these systems target interactive theorem provers such as Lean,
they rely on pretraining their neural networks on large corpora of existing
proofs, mathematical text, and code.
For instance, AlphaProof additionally uses a Gemini-based LLM to autoformalize approximately
one million natural-language problems into roughly 80 million formal Lean
statements as additional training data.
Our approach does not use pretraining and instead trains from scratch on individual protocols.

\paragraph{Machine Learning for Security Protocol Verification.}
The most closely related work is
SmartVerif~\cite{xiongSmartVerifPushLimit2020}, which integrates a Deep
Q-Network (DQN) into a modified version of Tamarin to select which
leaf node to expand in the proof search tree.
Each node is represented by a low-dimensional feature vector derived from the applied constraint reduction rule and the current step depth, although the paper does not specify how this vector is constructed.
SmartVerif's DQN greedily selects the highest-Q-value child at each step without any lookahead search, and it learns exclusively from a loop-detection heuristic based on Levenshtein distance between the string representation of the proof methods, assigning a uniform negative reward to paths identified as looping.

In our approach, we extend Tamarin with a reusable REST API instead of integrating the search directly into Tamarin.
We learn the encoding of a proof state through a transformer over tokenized proof methods, and our approach uses MCTS to proactively guide the search through learned and backed-up signals rather than relying on a fixed loop-detection heuristic.

Although the SmartVerif paper~\cite{xiongSmartVerifPushLimit2020} states that its artifacts are available, this no longer seems to be the case\footnote{They are
not accessible at the dropbox URL from the paper or through the authors' webpages. The v6 release of the corresponding ArXiv paper~\cite{smartverifarxivv6} contains a link to a github repo~\cite{smartverifgithub}, but this repository no longer exists. The latest v7 arXiv version~\cite{smartverifarxivv7} and journal version~\cite{smartverifarticle} do not provide any artifact links.}. Without access to SmartVerif's implementation and protocol formalizations, it is impossible to evaluate our approach on its models, or vice versa. Nevertheless, we chose several of our \emph{classical protocols} in~\Cref{sec:case_studies} inspired by SmartVerif's evaluation.

\section{Conclusion}
\label{sec:conclusion}
In this work, we have extended the Tamarin prover with an API for proof
search and used it to implement a new, reinforcement learning-guided proof search
in the style of AlphaZero~\cite{silverMasteringChessShogi2017}.
In contrast to Tamarin's current greedy search guided by internal or human-written
heuristics, our approach performs a graph search over the applicable constraint
solving rules and learns a policy and value function to guide this search. We have
evaluated our approach on 16 case studies, and we have found that it outperforms Tamarin's
heuristics in terms of the number of lemmas solved and the size of the proofs found. 
In particular, our approach performs well on the most complex, state-of-the-art models,
often achieving similar performance to human-written heuristics. Thus, our
work reduces the amount of human effort required to find proofs in Tamarin and
opens up new avenues for future research on proof search in Tamarin.


\bibliography{bibliography}

\clearpage

\appendix

\section{Additional Experimental Data}
\label{app:additional_experimental_data}

\paragraph{Hyperparameters.}
\Cref{tab:hyperparameters} lists the hyperparameters of the best configuration per protocol as reported in~\Cref{tab:lemma-completion}.
For WPA2, a second configuration ($\dagger$) solves one additional lemma not covered by the best configuration, as discussed in~\Cref{par:outperforming-baseline}.
Full configurations for reproducibility are in the artifact.

\begin{table*}[htbp]
  \centering
\caption{Hyperparameters of the best configuration per protocol. $B$: search budget, $B_{\text{inc}}$: budget increase factor, $S$: number of searches per lemma, $N$: expansion breadth, LR: learning rate, $\lambda$: heuristic influence, $\delta$: unvisited penalty, Temp: UCB temperature, $c_{\text{and}}$/$c_{\text{base}}$/$c_{\text{init}}$: UCB exploration constants, $r_{\text{time}}$: time penalty. $\dagger$: both runs solve the same number of lemmas, but each run solves one additional lemma that the other run does not solve.}
\label{tab:hyperparameters}
  \resizebox{\linewidth}{!}{%
  \begin{tabular}{@{}ll r r r r r r r r r r r r r@{}}
    \toprule
    Category & Protocol & Solved & $B$ & $B_{\text{inc}}$ & $S$ & $N$ & LR & $\lambda$ & $\delta$ & Temp & $c_{\text{and}}$ & $c_{\text{base}}$ & $c_{\text{init}}$ & $r_{\text{time}}$ \\
    \midrule
    \multirow{8}{*}{Classical Models} & FOO Eligibility & 3 & 1000 & 1.75 & 5 & 3 & 0.001 & 0.3 & 8 & 10 & 128 & 3200 & 0.0001 & 0.1 \\
     & KAS2 eCK & 1 & 1000 & 1.75 & 5 & 3 & 0.001 & 0.3 & 8 & 10 & 128 & 3200 & 0.0001 & 0.1 \\
     & NAXOS eCK & 1 & 1000 & 1.75 & 5 & 3 & 0.001 & 0.3 & 8 & 10 & 128 & 3200 & 0.0001 & 0.1 \\
     & NAXOS eCK PFS & 1 & 1000 & 1.75 & 5 & 3 & 0.001 & 0.3 & 8 & 10 & 128 & 3200 & 0.0001 & 0.1 \\
     & Tutorial & 4 & 1000 & 1.75 & 5 & 3 & 0.001 & 0.3 & 8 & 10 & 128 & 3200 & 0.0001 & 0.1 \\
     & UM PFS & 2 & 1000 & 1.75 & 5 & 3 & 0.001 & 0 & 8 & 10 & 128 & 3200 & 0.0001 & 0.1 \\
     & YubiKey & 2 & 1000 & 1.75 & 5 & 50 & 0.0001 & 0.3 & 8 & 10 & 64 & 3200 & 0.001 & 0.1 \\
     & YubiKey HSM & 4 & 1000 & 1.75 & 5 & 50 & 0.0001 & 0.3 & 8 & 10 & 64 & 3200 & 0.001 & 0.1 \\
    \midrule
    \multirow{4}{*}{Complex Models} & 5G Handover EPS N26 & 8 & 1000 & 1.75 & 5 & 3 & 0.0001 & 0.3 & 8 & 10 & 64 & 3200 & 0.001 & 0.1 \\
     & PKCS11 AEAD & 4 & 1500 & 1.75 & 20 & 3 & 0.001 & 0.3 & 8 & 10 & 128 & 3200 & 0.0001 & 0.1 \\
     & Signal & 5 & 12000 & 1.5 & 35 & 3 & 0.0001 & 0.5 & 8 & 50 & 64 & 50000 & 0.001 & 0.3 \\
     & Wireguard & 8 & 1500 & 1.5 & 15 & 3 & 0.0001 & 1 & 5000 & 200 & 64 & 10000 & 0.001 & 0.1 \\
    \midrule
    \multirow{5}{*}{\makecell[l]{State-of-the-Art\\Models}} & 5G AKA & 13 & 5000 & 1.5 & 15 & 3 & 0.0001 & 0.2 & 100 & 50 & 128 & 19652 & 0.0001 & 0.3 \\
     & 5G Handover XN & 9 & 5000 & 1.5 & 15 & 20 & 0.0001 & 3 & 50 & 10 & 0.01 & 3200 & 0.1 & 0.3 \\
     & SPDM & 37 & 2000 & 1.75 & 5 & 3 & 0.0001 & 0.5 & 8 & 10 & 128 & 3200 & 0.001 & 0.2 \\
     & WPA2$^\dagger$ & 73 & 1000 & 1.75 & 5 & 20 & 0.0001 & 0.3 & 8 & 10 & 128 & 3200 & 0.0001 & 0.1 \\
     & WPA2$^\dagger$ & 73 & 1500 & 1.75 & 5 & 3 & 0.0001 & 0 & 8 & 10 & 8 & 100 & 1 & 0.1 \\
    \bottomrule
  \end{tabular}}
\end{table*}

\paragraph{Proof Sizes.}
\Cref{tab:full-proofsize-overview} provides the full data on proof sizes for all lemmas solved by each approach, including those solved by only one approach. This complements the data in~\Cref{tab:proofsize-overview}, which is restricted to lemmas solved by both approaches.

\begin{table*}[htbp]
  \centering
  \caption{ Average proof size for the lemmas solved by the different approaches, i.e., RL, \z{c}, \z{s}, and Orig. Intersection columns are restricted to lemmas solved by both approaches. When \z{c} and \z{s} are aggregated into a single column, the smallest proof size per lemma is reported.}
  \label{tab:full-proofsize-overview}
  \resizebox{\linewidth}{!}{%
  \begin{tabular}{@{}llrrrrrrrr@{}}
    \toprule
    & & & \multicolumn{3}{c}{Tamarin} & \multicolumn{2}{c}{\z{c}/\z{s} $\cap$ RL} & \multicolumn{2}{c}{Orig.\ $\cap$ RL} \\
    \cmidrule(lr){4-6}\cmidrule(lr){7-8}\cmidrule(lr){9-10}
    Category & Protocol & \multicolumn{1}{c}{RL} & \multicolumn{1}{c}{\z{c}} & \multicolumn{1}{c}{\z{s}} & \multicolumn{1}{c}{Orig.} & \multicolumn{1}{c}{\z{c}/\z{s}} & \multicolumn{1}{c}{RL} & \multicolumn{1}{c}{Orig.} & \multicolumn{1}{c}{RL} \\
    \midrule
    \multirow{8}{*}{Classical Models} & FOO Eligibility & 20.3 & 102.3 & 21.0 & 21.0 & 21.0 & \textbf{20.3} & 21.0 & \textbf{20.3} \\
     & KAS2 eCK & 11.0 & 12.0 & 16.0 & 16.0 & 12.0 & \textbf{11.0} & 16.0 & \textbf{11.0} \\
     & NAXOS eCK & 117.0 & N/A & 134.0 & 134.0 & 134.0 & \textbf{117.0} & 134.0 & \textbf{117.0} \\
     & NAXOS eCK PFS & 8.0 & 13.0 & 13.0 & 13.0 & 13.0 & \textbf{8.0} & 13.0 & \textbf{8.0} \\
     & Tutorial & 8.2 & 13.2 & 9.0 & 9.0 & 9.0 & \textbf{8.2} & 9.0 & \textbf{8.2} \\
     & UM PFS & 8.0 & 8.0 & 8.0 & 8.0 & 8.0 & 8.0 & 8.0 & 8.0 \\
     & YubiKey & 29.0 & 12.0 & 12.0 & 12.0 & 12.0 & 12.0 & 12.0 & 12.0 \\
     & YubiKey HSM & 20.8 & 414.8 & 570.0 & 570.0 & 378.8 & \textbf{20.8} & 570.0 & \textbf{20.8} \\
    \midrule
    \multirow{4}{*}{Complex Models} & 5G Handover EPS N26 & 24.4 & 9532.7 & 9932.3 & 85.8 & 9531.7 & \textbf{17.3} & 85.8 & \textbf{24.4} \\
     & PKCS11 AEAD & 245.2 & 736.0 & 471.8 & 471.8 & 471.8 & \textbf{245.2} & 471.8 & \textbf{245.2} \\
     & Signal & 306.0 & 18.5 & 20.0 & 20.0 & 18.0 & \textbf{15.5} & 20.0 & 20.0 \\
     & Wireguard & 49.1 & 3816.2 & 92.0 & 47.8 & 77.4 & \textbf{49.1} & \textbf{47.8} & 49.1 \\
    \midrule
    \multirow{4}{*}{\makecell[l]{State-of-the-Art\\Models}} & 5G AKA & 194.0 & 27.4 & 1708.1 & 73.3 & 1679.0 & \textbf{148.8} & \textbf{73.3} & 118.6 \\
     & 5G Handover XN & 48.3 & 286.7 & 69.7 & 822.8 & 26.3 & \textbf{20.7} & 102.8 & \textbf{48.3} \\
     & SPDM & 70.8 & 270.9 & 85.9 & 104.8 & 94.9 & \textbf{64.6} & 106.1 & \textbf{70.8} \\
     & WPA2 & 52.0 & 1418.0 & 102.6 & 88.9 & 903.7 & \textbf{34.8} & 81.9 & \textbf{52.0} \\
    \bottomrule
  \end{tabular}}
\end{table*}

\paragraph{Times.}
\Cref{tab:protocol-time-minmax} gives a complete picture of the time required to solve all the lemmas that were in reach for the respective approach. We chose this per protocol aggregation instead of a per lemma listing since our approach trains and searches for proofs for all lemmas of a protocol concurrently. For a fair comparison with the baseline, we report the longest time it takes to solve a lemma across all lemmas solved, assuming maximal parallelization on the baseline.

Since our approach interleaves training and search, the reported times include both neural network updates and proof search via Tamarin.

On the classical models, all lemmas are solved within a couple of minutes.
On the complex models, the time ranges from a few minutes to about two hours:
5G Handover EPS N26 takes 3 minutes, WireGuard 5 minutes, PKCS11 AEAD 25 minutes, and Signal about 2 hours.
Notably, on 5G Handover EPS N26, our approach finishes faster than both of Tamarin's built-in heuristics individually:
3 minutes vs.\ 45 minutes (\z{c}) and 49 minutes (\z{s}).
On WireGuard, the fastest inbuilt configuration (\z{s}) is 23 minutes. Interestingly, \z{c} and \z{s} are complementary on WireGuard, such that the aggregate time of the best per-lemma time from either heuristic is 5 minutes 27 seconds, comparable to our 5 minutes 10 seconds. Our approach combines the best of both worlds into a single configuration that solves all lemmas in 5 minutes 10 seconds.
On the state-of-the-art models, training requires between 4.5 hours (5G Handover XN) and 20 hours (5G AKA), with SPDM at about 18 hours and WPA2 at about 13 hours.
While this is a significant computational cost, it remains well within a practical time frame and automates a significant part of the person-months of manual heuristic engineering that were necessary to achieve comparable results on these models.

\begin{table*}[htbp]
  \centering
  \caption{Wall-clock time to solve all lemmas of a protocol, assuming as much parallelization across lemmas as each approach allows.
  RL times include both training and proof search.
  Intersection columns pick the shortest time per lemma.
  Times are rounded to integer seconds (${\geq}10$s) or nearest minute (${\geq}1$h).}
  \label{tab:protocol-time-minmax}
  \resizebox{\linewidth}{!}{%
  \begin{tabular}{@{}llrrrrrrrr@{}}
    \toprule
    & & & \multicolumn{3}{c}{Tamarin} & \multicolumn{2}{c}{\z{c}/\z{s} $\cap$ RL} & \multicolumn{2}{c}{Orig.\ $\cap$ RL} \\
    \cmidrule(lr){4-6}\cmidrule(lr){7-8}\cmidrule(lr){9-10}
    Category & Protocol & \multicolumn{1}{c}{RL} & \multicolumn{1}{c}{\z{c}} & \multicolumn{1}{c}{\z{s}} & \multicolumn{1}{c}{Orig.} & \multicolumn{1}{c}{\z{c}/\z{s}} & \multicolumn{1}{c}{RL} & \multicolumn{1}{c}{Orig.} & \multicolumn{1}{c}{RL} \\
    \midrule
    \multirow{8}{*}{Classical Models} & FOO Eligibility & 31s & 4.9s & 1.3s & 1.3s & \textbf{1.3s} & 31s & \textbf{1.3s} & 31s \\
     & KAS2 eCK & 32s & 5.9s & 3.8s & 3.8s & \textbf{3.8s} & 32s & \textbf{3.8s} & 32s \\
     & NAXOS eCK & 3m 11s & N/A & 5.3s & 5.3s & \textbf{5.3s} & 3m 11s & \textbf{5.3s} & 3m 11s \\
     & NAXOS eCK PFS & 22s & 57m 22s & 5.0s & 5.0s & \textbf{5.0s} & 22s & \textbf{5.0s} & 22s \\
     & Tutorial & 18s & 0.5s & 0.4s & 0.4s & \textbf{0.4s} & 18s & \textbf{0.4s} & 18s \\
     & UM PFS & 18s & 3.9s & 0.8s & 0.8s & \textbf{0.8s} & 18s & \textbf{0.8s} & 18s \\
     & YubiKey & 2m 23s & 4.0s & 2.5s & 2.5s & \textbf{2.5s} & 31s & \textbf{2.5s} & 31s \\
     & YubiKey HSM & 49s & 48s & 34s & 34s & \textbf{26s} & 49s & \textbf{34s} & 49s \\
    \midrule
    \multirow{4}{*}{Complex Models} & 5G Handover EPS N26 & 3m 25s & 45m 6s & 49m 29s & 37s & 45m 6s & \textbf{48s} & \textbf{37s} & 3m 25s \\
     & PKCS11 AEAD & 25m 17s & 41s & 23s & 23s & \textbf{23s} & 25m 17s & \textbf{23s} & 25m 17s \\
     & Signal & 1h 50m & 1h 40m & 1m 12s & 48s & 1h 40m & \textbf{3m 22s} & \textbf{48s} & 1m 16s \\
     & Wireguard & 5m 10s & 1h 50m & 22m 37s & 1m 14s & 5m 27s & \textbf{5m 10s} & \textbf{1m 14s} & 5m 10s \\
    \midrule
    \multirow{4}{*}{\makecell[l]{State-of-the-Art\\Models}} & 5G AKA & 19h 48m & 29m 27s & 11m 32s & 17m & \textbf{11m 32s} & 2h 27m & \textbf{17m} & 19h 48m \\
     & 5G Handover XN & 4h 36m & 1m 8s & 42s & 1h 28m & \textbf{39s} & 1m 59s & \textbf{3m 22s} & 4h 36m \\
     & SPDM & 17h 48m & 3h 42m & 3h 6m & 49m 21s & \textbf{21m 23s} & 17h 48m & \textbf{17m 49s} & 17h 48m \\
     & WPA2 & 12h 55m & 1h 8m & 12m 23s & 5m 58s & \textbf{1h 8m} & 1h 52m & \textbf{5m 58s} & 12h 55m \\
    \bottomrule
  \end{tabular}}
\end{table*}

\section{Original Results Reproduction}
\label{app:original_results}
We fail to prove all lemmas proven in the original models of the YubiKey, Signal, 5G AKA, SPDM, and WPA2 case studies, even though we used timeouts of up to four hours per lemma, which is significantly longer than the timeouts used in the original papers.
We discuss potential reasons for this failure below.

The YubiKey paper~\cite{kunnemann2012yubisecure} reports that all lemmas were solved, but this required finding \emph{typing invariants} via Tamarin's \emph{source} lemmas.
However, the models did not contain such lemmas, and Tamarin did not report any \emph{partial deconstructions}, which are usually a sign of such invariants being necessary.
Additionally, we tried running the models using Tamarin's \emph{auto-sources} feature, which automatically adds sources lemmas, but this did not help either.
We suspect that either the original proofs were done manually or that we were not able to find the proofs
automatically because of differences in Tamarin versions.

For the Signal model from~\cite{linker2025looping}, we only solve 1 out of 5 lemmas with the authors heuristics, while the paper reports that all lemmas were solved.
\cite{linker2025looping} extends Tamarin with cyclic induction and uses the Signal protocol model to compare Tamarin's performance under trace induction against their cyclic induction extension.
Since we solve all lemmas using our approach, we are sure that the model does not \emph{need} the cyclic induction extension (our Tamarin version does not have the cyclic induction extension). 
Again, we suspect our failure to reproduce the original results are due to different Tamarin versions: Both our work and the original work modify Tamarin and
we forked different releases of the main repository.

For the 5G AKA and the SPDM model, we fail to reproduce the proof of a single lemma. We ran these lemmas again with a four hour timeout instead of two, but we still failed to prove them.
Since we solve the other lemmas for these models, the difference in Tamarin versions is less likely to be the cause of this failure, but it still might be. Another reason for our failure might be that we are not giving Tamarin enough computational resources. While our hardware is powerful, we are proving multiple lemmas concurrently, and we are not the only ones using the compute infrastructure.
As a consequence, there is some non-determinism in the actual available computational resources.

For the WPA2 model, we fail to reproduce the proofs for 10 existential lemmas. Again, this is likely due to the different Tamarin versions or insufficient computational resources.

\end{document}